\documentclass[12pt]{elsart3p}
\usepackage{graphicx,amssymb}
\usepackage{color}
\begin{document} 
\begin{frontmatter}
\title{High $p_T$ Hadron Correlation and No Correlation}
\author{Rudolph C. Hwa}
\address {Institute of Theoretical Science and Department of Physics\\ University of Oregon, Eugene, OR 97403-5203, USA}
\ead{hwa@uoregon.edu}
\begin{abstract}
An overview of the correlation among high-$p_T$ hadrons produced in heavy-ion collisions is presented. Emphasis is placed on the physical processes that can quantitatively account for the data on correlations in $p_T, \eta$ and $\phi$ on the near and far sides. Predictions are made on processes that have no observable correlations for hadrons produced at intermediate and higher $p_T$ at RHIC and LHC.

\end{abstract}
\end{frontmatter}

\section{Introduction}

The conventional understanding about heavy-ion collisions is that when there is a hard scattering of partons, a high-$p_T$ jet is created and one studies the correlation of hadrons either within the same jet or on opposite sides.  The unconventional scenario is that it is possible to detect a high-$p_T$ hadron without an identifiable high-$p_T$ jet, and thus there would be no correlation of particles above the background.  Both possibilities will be covered in this review.  The former is important because that is how we learn about the properties of the hot and dense medium through jet quenching.  The latter cautions us from regarding every high-$p_T$ particle as being a part of a jet.  Whether correlation exists or not depends on the hadron species, the medium from which they are produced, the kinematical region where they are detected, etc.  But most important of all is the dynamical properties of the partons just before and during hadronization.  Thus both possibilities are essential to a complete understanding of the medium created in heavy-ion collisions.

For jet correlation we consider near- and far-side correlations in  
$p_T$, $\eta$ and $\phi$, as well as auto-correlation.  For no jet correlation there are three areas where the phenomenon is expected to occur:  $\phi$ and $\Omega$ production, $p$ and $\pi$ production at large $\eta$, and at large $p_T$ at LHC. Only ideas and results will be presented here, quantitative details being referred to the original papers.

\section{Hadron Correlation in Jets}

Since most of the existing data on jet correlation are for $p_T^{\rm trig}< 6$ GeV/c, and for $p_T^{\rm assoc}$ even lower, fragmentation is not a reliable mechanism for hadronization at those intermediate momentum ranges \cite{rch}.  
Most of the quantitative calculations that can reproduce the data are based on the recombination model (RM).  We give here a brief overview of such studies.

\subsection{Correlation in $p_T$ on the near side}

The $p_T$ distribution of associated particles in the same jet that contains the trigger particle has been calculated in the RM, where the thermal-shower parton recombination plays the dominant role \cite{ht}.  The results are shown in Fig.\ 1 for both (a) $d + Au$ and (b) $Au + Au$ collisions at 200 GeV at various centralities.  Evidently, there is little dependence on centrality in $d + Au$ collisions, but significant dependence in $Au + Au$ collisions.  Those results are in accord with the data.  Fig.\ 2 shows the recent data from STAR presented at HP06 \cite{jb}.  The solid line for $\pi \pi$ correlation \cite{ht} agrees well with the square points for $hh$ correlation in $Au + Au$ collisions.

\begin{figure}
\begin{center}
\includegraphics*[width=7cm]{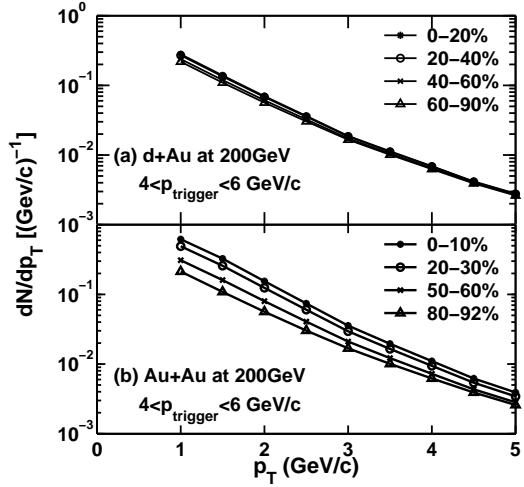}
\end{center}

\caption{Associated particle distribution in the RM \cite{ht}}
\label{fig:fig1}
\end{figure}

\begin{figure}
\begin{center}

\hspace*{-2mm}
\includegraphics*[width=7cm]{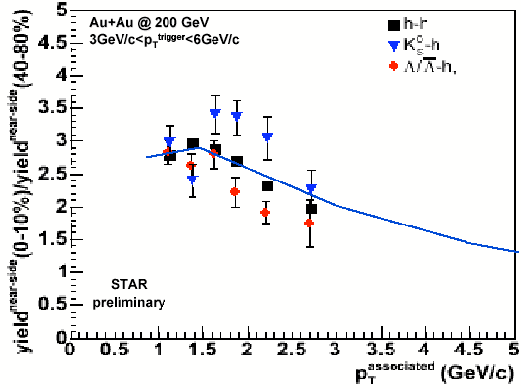}
\end{center}

\caption{Central-to-peripheral ratio of particles on the near side \cite{jb}. Solid line is from Fig.\ 1.}
\label{fig:fig2}
\end{figure}

\subsection{Correlation in $p_T$ on the away side}

On the away side there are data for $4 < p_T^{\rm trig}< 6$ GeV/c and $0.5 < p_T^{\rm assoc}<4$ GeV/c from STAR \cite{ja} as shown in Fig.\ 3.  From the ratio of $AA/pp$ shown in (d) it is clear that there is suppression at $p_T^{\rm assoc}> 2$ GeV/c but enhancement at $<2 $ GeV/c.  That is quite understandable as the consequence of jet quenching in the higher $p_T^{\rm assoc}$  region and soft-parton regeneration in the lower region.  So far there has been no quantitative theoretical calculation that reproduces these properties of the data.

\begin{figure}
\begin{center}
\hspace*{-15mm}
\includegraphics*[width=5cm]{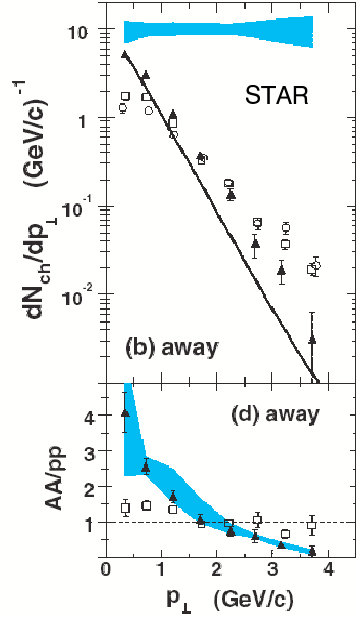}
\end{center}
 
\caption{Away-side correlation for $4<p_T^{\rm trig}<6$ GeV/c and $0.5<p_T^{\rm assoc}<4$ GeV/c \cite{ja}.}
\label{fig:fig3}
\end{figure}

\vspace{1mm}

For jet tomography it is necessary to go to higher trigger momentum, where the medium modified fragmentation function can be more reliably applied \cite{gw}.  STAR has data with $p_T^{\rm trig}$ in the 8-15 GeV/c range \cite{ja2}.  The away-side jets show diminishing peak heights as $p_T^{\rm assoc}$ is increased.  To quantify that behavior the trigger-normalized fragmentation function $D(z_T)$ is studied, where $z_T$ is the trigger-normalized momentum fraction $p_T^{\rm assoc}/p_T^{\rm trig}$, proposed by Wang \cite{xw}.  Shown in Fig.\ 4 
is $D(z_T)$ vs $z_T$ on the away side for $z_T>0.35$ \cite{ja2}.  The straight lines are drawn to indicate the universality of the $z_T$ dependence among $d+Au$ and $Au+Au$ collisions, claimed by STAR, if the points at $z_T=0.35$ are ignored.  It seems that if the lowest point for $d + Au$ min bias is also ignored, the best fit for that collision system would be a straight line with a lower slope, in comparison with which the data points for the $Au+Au$ system would fall off faster with $z_T$.  To focus on the ratio $D_{AuAu}/D_{dAu}$, as shown in the lower panel of Fig.\ 4, may be misleading, since all three $D(z_T)$ deviate strongly from linear extrapolations at $z_T = 0.35$, while their ratios reveal no such deviation.    Our point here is to register an alert for the possible transition of the hadronization mechanism between $z_T>0.6$ and $z_T<0.5$ regions.  The higher $z_T$ region is dominated by fragmentation and exhibits universality, while the lower $z_T$ region shows signs of enhancement due to the contribution from recombination.  We suggest that the $p/\pi$ ratio should be measured on the away side.  Its dependence on $z_T$ would reveal the proportion of the two mechanisms:  the larger $R_{p/\pi}$ is, the more is the participation of recombination, which is a property of the RM that has been established in the study of single-particle distributions \cite{hy,gr,fr}.

\begin{figure}
\begin{center}
\vspace*{-15mm}
\hspace*{-50mm}
\includegraphics*[width=16.5cm]{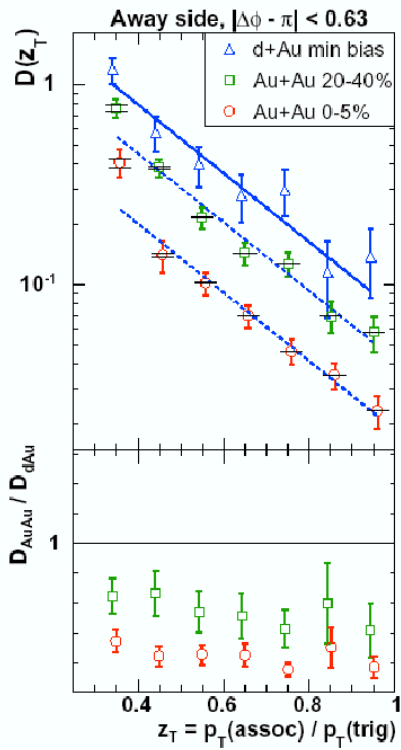}
\end{center}

\vspace{-18mm}
\caption{Trigger-normalized fragmentation functions \cite{ja2}.}
\label{fig:fig4}
\end{figure}

\vspace{1mm}
\subsection{Correlation in $\Delta\eta$ and $\Delta\phi$ on the near side}

Turning next to the correlation in $\eta$ and $\phi$ on the near side, we recall the characteristic of the STAR data \cite{ja} that show a peak in $\Delta \eta$ sitting on top of a pedestal that stretches over the interval $|\Delta\eta| \lesssim 1.2$.  The pedestal may well be influenced by the effect of longitudinal expansion \cite{arm}.  In the RM we have been able to reproduce both the peak and the pedestal, as shown in Fig.\ 5 \cite{ch}, by relating the thermalization of the soft partons generated by the energy loss of the hard parton to the pedestal with $\Delta T \approx 15$ MeV.  The peak is due to the recombination of the secondary shower parton with the thermal parton, the primary shower parton being responsible for the trigger particle.  The prediction for the distribution in $\Delta \phi$ agrees well with the data \cite{ch}.  At higher trigger momentum the hard partons originate closer to the surface, so less energy is lost to the medium, and the pedestal phenomenon goes away.

\begin{figure}
\begin{center}
\hspace*{-7mm}
\includegraphics*[width=6cm]{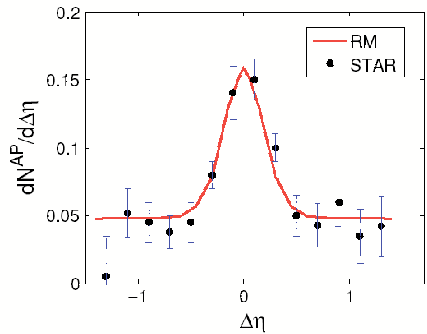}
\end{center}
 
\caption{Near-side peak and pedestal with data from \cite{ja} and solid line from \cite{ch}.}
\label{fig:fig5}
\end{figure}

\vspace{1mm}
\subsection{Correlation in $\Delta\phi$ on the away side}
Compared to the near-side correlation, there are far more theoretical studies done on the $\Delta \phi$ distribution on the away side, since that is where the properties of hard parton propagating through a dense medium are revealed.  Various suggestions have been made, such as Mach cone, Cherenkov gluons and radiation, color wake, jet quenching and fragmentation \cite{cs}-\cite{vit}.  Most of the suggestions describe the collective response of the medium to the passage of a hard parton.  Such a response, especially in the case of Mach cone, would result in enhancement of hadron production on both side of the recoil jet in every triggered event so that the $\Delta \phi$  distribution would exhibit a double bump structure around $\Delta \phi = \pi$, as has been observed for $2.5 < p_T^{\rm trig} < 4$ and  $1 < p_T^{\rm assoc} < 2.5$ GeV/c \cite{jia}.  

An alternative approach to the problem is presented by Chiu at this meeting \cite{ch2}.  A parton multiple scattering model is constructed to reproduce the dip-bump structure in the data of  \cite{jia} by allowing the recoil parton to undergo multiple scattering as it traverses the dense and expanding medium with random scattering angle in the forward cone at each step.  The parton loses energy at each step so it may be absorbed by the medium if the trajectory has a long path length.  The exit tracks have the characteristics that they bend persistently in one direction and leave the medium in a few steps.  While each track in this semi-classical description may not be realistic, the general features after averaging over all tracks present a sensible modeling of deflected jets.  The main phenomenological feature is that there is at most one deflected jet per triggered event, but the event-averaged $\Delta \phi$ distribution exhibits the symmetrical dip-bump structure around $\varphi=\Delta\phi-\pi=0$, as shown in Fig.\ 6.  The dashed line is the pedestal due to thermalized soft partons generated by energy loss.  The solid line above the pedestal describes the hadron distribution due to the exit partons.  When the event generator is applied to the case with higher trigger momentum in the range $8  < p_T^{\rm trig} < 15$ GeV/c \cite{ja2}, there is only one peak on the away side at  $\Delta \phi = \pi$, with its height and width in agreement with the data shown in Fig.\ 7 for (a) $4 < p_T^{\rm assoc} < 6$ GeV/c and (b) $ p_T^{\rm assoc} > 6$ GeV/c.

\begin{figure}
\begin{center}
\vspace*{-5mm}
\hspace*{-5mm}
\includegraphics*[width=8cm]{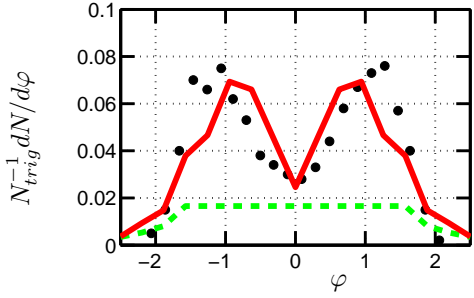}
\end{center}
\vspace*{-5mm}
\caption{$\Delta\phi$ distribution on the away side. Data are from \cite{jia}. Dashed line is the calculated pedestal, and solid line includes the deflected jets \cite{ch2}.}
\label{fig:fig6}
\end{figure}

\begin{figure}
\begin{center}
\hspace*{-5mm}
\includegraphics*[width=8cm]{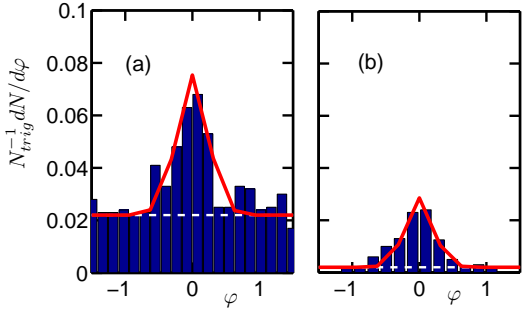}
\end{center}
\vspace*{-5mm}
\caption{$\Delta\phi$ distribution on the away side for $8<p_T^{\rm trig}<15$ GeV/c and (a) $4<p_T^{\rm assoc}<6$ GeV/c and (b) $p_T^{\rm assoc}>6$ GeV/c.}
\label{fig:fig7}
\end{figure}

\subsection{Autocorrelation}

The advantage of studying autocorrelation is that two particles are treated on the same footing and that a larger set of angular relationships between two shower partons in the same jet is included in the study of correlation compared to the analysis in which  one particle  is treated as a trigger and the other as the associated particle. The usual correlation function (cumulant), $C_2(1,2)=\rho_2(1,2)-\rho_1(1)\rho_1(2)$, is defined in terms of the momenta $\vec p_1$ and $\vec p_2$ of the two particles, which in turn can be expressed in terms of angular differences and sums $\theta_{\pm}, \phi_{\pm}$, ignoring their magnitudes if the emphasis is on $\Delta\eta$ and $\Delta\phi$ analysis. Autocorrelation $A(\eta_-,\phi_-)$ is the integral of $C_2$ over $\theta_+$ and $\phi_+$. No ambiguous subtraction procedure is used. Data from STAR have been analyzed yielding a wealth of information on autocorrelation \cite{ja3}.

\begin{figure}
\begin{center}
\hspace*{-5mm}
\includegraphics*[width=6cm]{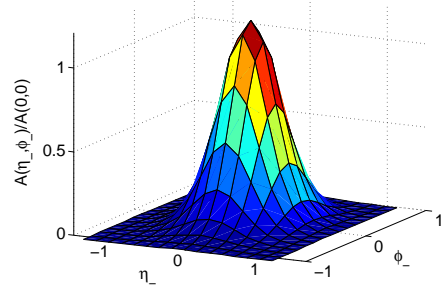}
\end{center}

\vspace{-10mm}
\caption{Autocorrelation with $\eta_-=\Delta\eta$ and $\phi_-=\Delta\phi$ \cite{ch3}.}
\label{fig:fig8}
\end{figure}

In the RM autocorrelation is studied by considering two shower partons in a jet oriented in all possible directions in the jet cone. They hadronize by recombination with thermal partons for $p_T$ in the intermediate region \cite{ch3}. In Fig.\ 8 is shown the normalized autocorrelation function  for the width of the jet cone being $\sigma=0.2$. Of course, other values of $\sigma$ can be considered, resulting in different shapes of $A(\eta_-,\phi_-)$. That is where one envisions the possibility of extracting the width of the jet cone from the data by comparing the structure in Fig.\ 8, or its projections to either $\eta_-$ or $\phi_-$, with similar experimental plots. So far the data on autocorrelation is for $p_T$ too low to be used meaningfully in the above sense, since there is no significant contribution of shower partons at low $p_T$.

\section{High $p_T$ Hadrons Without Correlation}

Before we go into specific problems that depart from the conventional situation where correlation always accompanies hadrons produced at high $p_T$, let us discuss in general the physics of hadron correlation at high $p_T$.  As we have seen in the previous section, when an event is triggered by a high $p_T$ particle, it is usually due to a jet that supplies not only the trigger particle, but also other particles in the same jet, as well as a recoil jet that sends particles in roughly the opposite direction.

Thus the detection of high-$p_T$ particles is a necessary consequence of high-$p_T$ partons, but is it sufficient?  As for how high is ``high $p_T$'',  we first consider the case where $p_T$ is around 4 GeV/c at RHIC, and later between 10 and 20 GeV/c at LHC.

It has been shown in \cite{hy} that for single-particle inclusive distribution in the $3 < p_T < 8$ GeV/c range the dominant mechanism of pion production in $Au Au$ collision at 200 GeV is the recombination of shower partons with thermal partons.  That is the way that the medium effect is taken into account in the RM for the production of high-$p_T$ pions and protons.  The recombination of thermal partons among themselves is important at lower $p_T$.  The shower partons are generated by high $p_T$ partons in hard scattering, so indeed the pions and protons produced at high $p_T$ are the necessary consequences of high-$p_T$ partons.  But what if  the hard scattering processes are suppressed for dynamical and kinematical reasons?  Then in the absence of shower partons the thermal partons would emerge as the only partonic source for hadronization even for particles at $p_T \sim 4$ GeV/c or higher.  Although the consideration here is only for single-particle distribution, it is clear that the nature of correlation would be drastically altered when the suppression of hard scattering removes the dynamical origin of high-$p_T$ correlation of the usual kind.  Indeed, one should not expect any correlation if the particles produced are of thermal origin.

The scenario described above turns out to show up in two cases:  $\phi$ and $\Omega$ production at mid-rapidity, and $\pi$ and $p$ production at forward rapidity.  Another special case involves not so much the lack of hard scattering, but too many of them at LHC.  In that case correlation cannot stand out above the background, so the detection of dynamical correlation should yield null result.

\begin{figure}
\begin{center}
\hspace*{-5mm}
\includegraphics*[width=5.5cm]{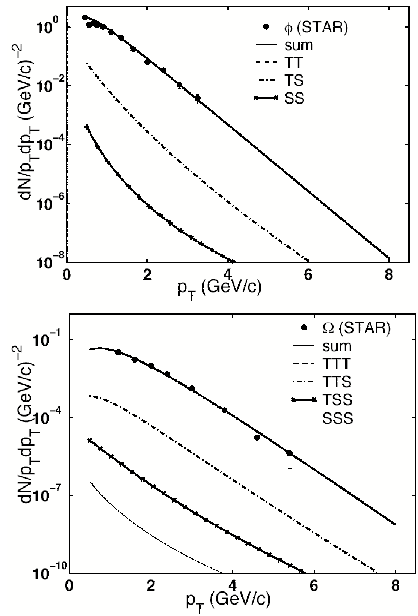}
\end{center}

\caption{$\phi$ and $\Omega$ distributions in the RM \cite{hy2}.}
\label{fig:fig9}
\end{figure}

\subsection{ $\phi$ and $\Omega$  production}

Since  $\phi$ and $\Omega$  are made of strange quarks only, if they are to be produced by hard scattering, the relevant processes involve either (a) a high-$p_T$ non-strange parton that creates an $s$ shower parton, or (b) a high-$p_T$ $s$ parton by direct scattering.  In the former case the $q \rightarrow s$ showering process is suppressed, while in the latter case the direct production of high-$p_T$ $s$ quark is suppressed.  In either case the distribution of $s$ shower parton is weak, so thermal-shower recombination is not important in the intermediate $p_T$ region.  That leaves only the thermal $s$ partons to recombine to form  $\phi$ and $\Omega$ .  In Fig.\ 9 are shown  the $p_T$ distributions of  $\phi$ and $\Omega$ for all combinations of thermal $(T)$ and shower $(S)$ recombination \cite{hy2}, assuming $\delta$-function for the recombination functions.  Notice how much lower the contributions involving $S$ are compared to the dashed lines that are nearly completely covered by the solid lines that represent the sums.  Thus in the intermediate $p_T$ region those distributions are approximately exponential, which reveals he thermal nature of the recombining partons.  The agreement with the existing data in Feb 2006 is excellent.

\begin{figure}
\begin{center}
\hspace*{-5mm}
\includegraphics*[width=6.5cm]{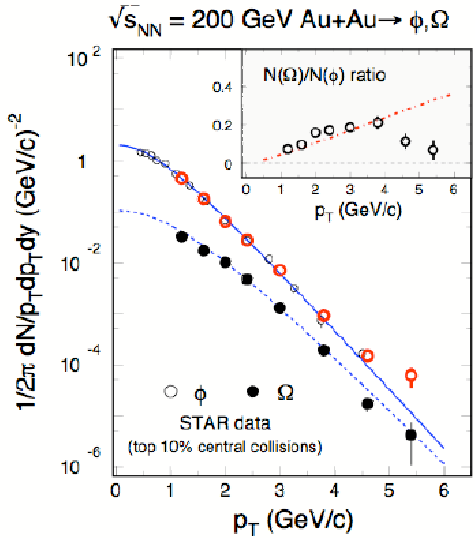}
\end{center}
\caption{$\phi$ and $\Omega$ distributions from STAR \cite{nx}.}
\label{fig:fig10}
\end{figure}

Subsequently, STAR has extended those data to higher $p_T$, as shown in Fig.\ 10 \cite{nx}.  It is evident that the $\phi$ points at high-$p_T$ deviates from the straight-line extrapolation of \cite{hy2}, although the $\Omega$ points seem to retain the exponential behavior out to 5.5 GeV/c.  The theoretical reason for the up-bending of $\phi$ is currently under investigation.  The ratio $R_{\Omega/\phi}$ being linearly rising up to $p_T \approx 4$ GeV/c is nevertheless very remarkable.  
The linearity of the line in the inset is due to dimensional differences in the equations for 2 and 3-quark recombination of thermal partons, but the significance is in the thermal nature of the partons. Thus we find
 that the $s$ quarks are thermal out to 2 GeV/c, a range of validity far wider than one normally associates with statistical distribution for $T \approx 0.32$ GeV/c.  One may infer from this result that the medium has the characteristics of a quark-gluon plasma, which are:  (a) strange quarks are enhanced, and (b) they are thermalized.

Now let us return to the issue about correlation.  Since Figs.\ 9 and 10 indicate that the spectrum of $\Omega$ is exponential out to $p_T \approx 6$ GeV/c, a property that is interpreted in the RM as being of thermal origin without the involvement of hard partons, then there can be no hadrons correlated with the observed $\Omega$ up to 6 GeV/c that are not in the background \cite{hy2}.  
This prediction is currently under active investigation by STAR.

\subsection{Forward production at any $p_T$}

The production of hadrons at large $\eta$ is another situation where hard scattering is suppressed.  If the hadronization process is fragmentation, then a hard parton must first be scattered to nearly the kinematical boundary in order for it to fragment to a hadron at large $\eta$.  But since such hard scattering must vanish at the kinematical boundary, it is highly suppressed just inside the boundary.  On the other hand, hadronization by recombination involves the additivity of parton momenta so nothing needs to vanish; indeed, the distribution of produced hadron can vary smoothly across the kinematical boundary separating the fragmentation region (FR, $x_F < 1$) and the trans-fragmentation region (TFR, $x_F > 1$) \cite{hy3}.  For example, three quarks at $x \approx 0.4$ can recombine and form a baryon at $x_F \approx 1.2$.  That is not possible in $pA$ collisions, since all three forward moving quarks must come from the projectile proton.  But in $AA$ collisions the three quarks can originate from different nucleons in the projectile nucleus, so each having $x \approx 0.4$ is not forbidden.

There are two complications in forward production at moderate $p_T$ in $AA$ collisions that are absent at high $p_T$ and mid-rapidity.  One is the degradation of longitudinal momentum that has been referred to as ``baryon stopping'' in an archaic language.  The other is the regeneration of soft partons due to energy loss by the leading partons.  The former has been studied in detail in $pA$ collisions \cite{hy4} and then in $AA$ collisions \cite{hy3}.  The latter is considered in \cite{hy5}.  The regeneration of soft partons does not matter too much to the production rate of forward proton, for which valence quarks dominate, but its effect on pion production is significant.  Since these are non-perturbative processes, the calculations are model dependent and data guided.

The most recent data that are relevant are from BRAHMS for $\eta = 3.2$ at $\sqrt{s} = 62.4$ GeV \cite{ars}, where the $p_T$ distribution is given.  With both $\eta$ and $p_T$ known, it is possible to determine $x_F$.  Unfortunately, just as $x_F$ approaches 1, the analysis of the $p_T$ distribution is stopped, although events have been detected with some particles having $x_F >1$.  Nevertheless, for $p_T$ up to 2.2 GeV/c, the charged particle spectrum shows exponential behavior \cite{ars}; it corresponds to $x_F$ in the mid-fragmentation region between 0.4 and 0.9.

\begin{figure}
\begin{center}
\vspace*{-2mm}
\hspace*{-5mm}
\includegraphics*[width=7cm]{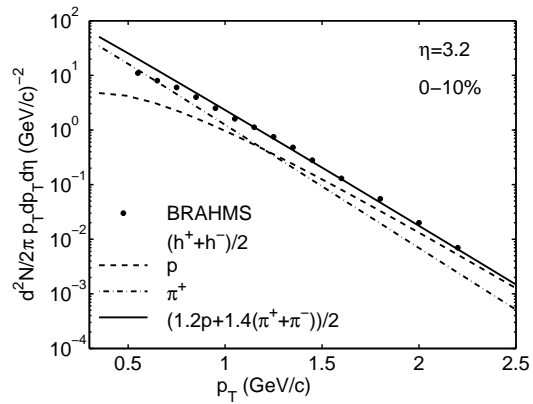}
\end{center}
\vspace*{-1mm}
\caption{$p_T$ distribution in  forward production. Data from \cite{ars}; lines from \cite{hy5}.}
\label{fig:fig11}
\end{figure}

Our study in the RM has been able to reproduce both the $p_T$ distribution and the $p/\pi$ ratio \cite{hy5}.  The results are shown in Figs.\ 11 and 12.  There are no shower partons contributing because hard scattering is suppressed in the FR.  The solid line in Fig.\ 10  for $[1.2p+1.4(\pi^++\pi^-)]/2$ that includes $\bar p$ and $K^{\pm}$ approximates the average charged particles, based on $p$ and $\pi$ distributions that are determined by the recombination of thermal partons only,  including the regenerated soft partons. The agreement with data is good. Note that the $p/\pi$  ratio in Fig.\ 12 reaches as high as 2 at $p_T \sim 2$ GeV/c.  That is impossible by parton fragmentation, but in recombination it is actually difficult to keep it as low as 2 unless there is substantial regeneration of antiquarks.  The reason is that valence quarks dominate in the FR so the formation of proton by recombination is much easier than to form pion due to the scarcity of antiquarks whose density depends sensitively on the regeneration effect.  The degradation factor $\kappa$ has been adjusted to reproduce the data \cite{hy5,mm}.

\begin{figure}
\begin{center}
\vspace*{-1mm}
\hspace*{-5mm}
\includegraphics*[width=6cm]{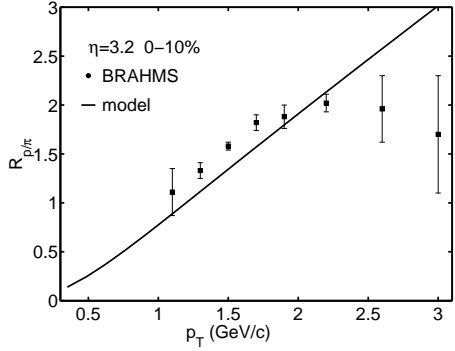}
\end{center}
\vspace*{-1mm}
\caption{$p/\pi$ ratio at $\eta=3.2$ and $\sqrt s=62.4$ GeV. Data from \cite{mm}; line from \cite{hy5}.}
\label{fig:fig12}
\end{figure}

Since no shower partons are involved in the above calculations, no hard partons are needed for $p_T > 2$ GeV/c, neither are they available due to the suppression of hard scattering in the forward region.  Thus there are no jets, and one should not expect to observe any hadrons correlated at any $p_T$.

\subsection{Hadron production at high $p_T$ at LHC}

At LHC it is predicted that there can be a large number of jets, yet no jet structure is likely to be observed \cite{hy6}.  That is very peculiar, and therefore interesting to verify.  The reasoning is very nearly the opposite of those in the previous two subsections.  It is  anticipated that the jet multiplicity is high at LHC, roughly 100 in $|\eta|< 0.5$.  Since the shower partons generated by a high-$p_T$ parton are distributed in a jet cone of non-negligible size, the overlap of jet cones of neighboring jets is unavoidable when the density of jets is high.  The probability of that overlap may be difficult to calculate, but given two or three colinear partons arising from two adjacent jets, their recombination to form pion and proton is easy to calculate in the RM.  Thus if we allow the overlap probability $\Gamma$ to vary over a wide range, from $10^{-1}$ to $10^{-4}$, we can determine the range of single-particle distribution of $\pi$ and $p$ in the region $10 < p_T < 20$ GeV/c, high enough to leave out the thermal partons, and low enough to have high jet density.

In \cite{hy6} such a calculation has been done with the results shown in Fig.\ 13.  The parameter $\xi$ refers to the suppression factor of hard partons emerging from the super-dense medium, and is related to $R_{AA}$.  At RHIC it is $\xi = 0.07$ \cite{hy}.  Since it is unknown at LHC, two reasonable values are used in Fig.\ 13.  The more important dependence of the distributions is on $\Gamma$.  The pion distributions in the left panel reach their limiting behavior of no 2-jet recombination at $\Gamma = 10^{-4}$, but the proton distributions on the right continue to decrease in magnitude at small $\Gamma$.  The interpretation is that proton is so much more abundantly produced because of 3 quark recombination that even at $\Gamma = 10^{-4}$ the 2-jet contribution is larger than the 1-jet contribution, which is for $\Gamma = 0$ (not shown).  Since it is expected that $\Gamma$ decreases with $p_T$, a form $\Gamma(p_T) \sim p_T^{-7}$ is used to see how the distributions approach the 1-jet limit at high $p_T$.  That is shown by the heavy lines in Fig.\ 13.  For $\pi$ they approach the limiting curves before $p_T$ reaches 20 GeV/c, but for $p$ they do not.  Since for $p_T > 20$ GeV/c the average jet multiplicity is less than 2 in $|\eta| < 0.5$, $\Gamma (p_T)$ is expected to decrease much faster than $p_T^{-7}$, so the heavy line for $p$ should rapidly join the $\Gamma = 0$ line beyond.

\begin{figure}
\begin{center}
\hspace*{-5mm}
\includegraphics*[width=85mm]{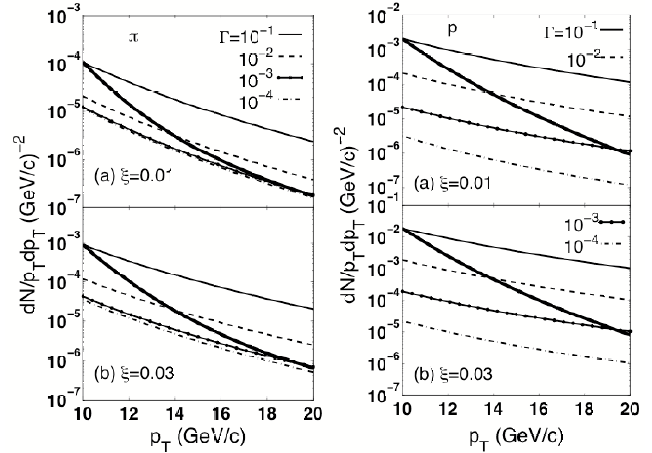}
\end{center}
\vspace*{-1mm}
\caption{$\pi$ and $p$ distributions at LHC for various 2-jet overlap probability $\Gamma$. Heavy line represents $\Gamma\sim p_T^{-7}$ \cite{hy6}.}
\label{fig:fig13}
\end{figure}

\begin{figure}
\begin{center}
\vspace*{-2mm}
\hspace*{-5mm}
\includegraphics*[width=5cm]{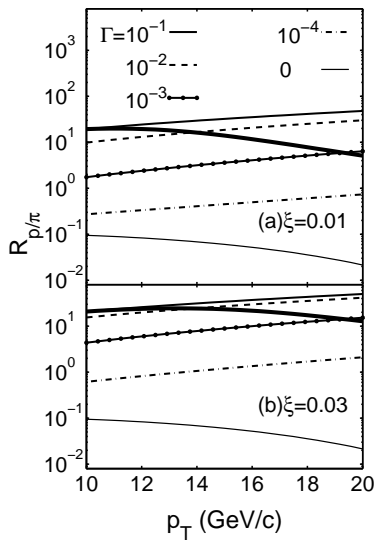}
\end{center}
\vspace*{-1mm}
\caption{$p/\pi$ ratio at LHC. Symbols as in Fig.\ 13.}
\label{fig:fig14}
\end{figure}

The important phenomenological consequence of the above results is the ratio $R_{p/\pi}$ that is insensitive to $\xi$.  That ratio is shown in Fig.\ 14.   For fixed $\Gamma$ the curves increase with $p_T$, reflecting the dominance of 3-quark recombination.  For  $\Gamma(p_T) \sim p_T^{-7}$, the heavy lines show that  $R_{p/\pi}$ decreases from 20 to 5, but is significantly higher than that of the one-jet case, shown by the light solid lines in the same figure.  Such a large $p/\pi$ ratio at such a high region of $p_T$ is a result that differs spectacularly from what one expects from fragmentation in the conventional approach to high-$p_T$ physics.

In addition to the surprisingly large $p/\pi$ ratio, there is also the unusual correlation characteristics. At RHIC, where there is a jet, there are correlated hadrons. But at LHC so many jets are produced, they are part of the background. Since hadrons  in the $10<p_T<20$ GeV/c range are not the fragments of any super-hard parton at much higher $p_T$, but are the recombination products of semihard partons at lower $p_T$ that are numerous in every event, treating a detected hadron as a trigger does not eliminate any other hadrons formed by other jets in the same event. That is, there are many associated particles, all of which are part of the background. Thus it is not possible to find any correlated particles distinguishable from the background \cite{hy6}. That property is independent of $\Delta\phi$, since recoil partons are weaker than the ambient jets produced near the surface on the far side. Data on these predictions can provide crucial guidance to the understanding of high-density medium whose emitted particles at high $p_T$ has usually been described in a framework that may have to be modified.

I am grateful to  C.\ B.\ Chiu, Z.\ Tan and C.\ B.\ Yang for their collaboration. This work was supported, in part,  by the U.\ S.\ Department of Energy under Grant No. DE-FG02-96ER40972.

\end{document}